%% file: main.tex
\documentclass[pmlr]{jmlr_1}

\usepackage{longtable}

\usepackage{booktabs}
\usepackage[load-configurations=version-1]{siunitx} 

\usepackage{amssymb}

\usepackage{graphicx}
\usepackage{color}
\usepackage{xcolor}
\usepackage{amssymb}
\usepackage{hyperref}
\usepackage{wrapfig}
\usepackage{caption}
\usepackage{booktabs}
\usepackage{amsmath,amsfonts,bm, bbm}
\usepackage[edges]{forest}
\usepackage{url}
\usepackage{tablefootnote}
\makeatletter
\newcommand\footnoteref[1]{\protected@xdef\@thefnmark{\ref{#1}}\@footnotemark}
\makeatother

\definecolor{foldercolor}{RGB}{200,200,200}

\tikzset{pics/folder/.style={code={%
    \node[inner sep=0pt, minimum size=#1](-foldericon){};
    \node[folder style, inner sep=0pt, minimum width=0.3*#1, minimum height=0.6*#1, above right, xshift=0.05*#1] at (-foldericon.west){};
    \node[folder style, inner sep=0pt, minimum size=#1] at (-foldericon.center){};}
    },
    pics/folder/.default={20pt},
    folder style/.style={draw=foldercolor!80!black,top color=foldercolor!40,bottom color=foldercolor}
}

\forestset{is file/.style={edge path'/.expanded={%
        ([xshift=\forestregister{folder indent}]!u.parent anchor) |- (.child anchor)},
        inner sep=1pt},
    this folder size/.style={edge path'/.expanded={%
        ([xshift=\forestregister{folder indent}]!u.parent anchor) |- (.child anchor) pic[solid]{folder=#1}}, inner xsep=0.75*#1},
    folder tree indent/.style={before computing xy={l=#1}},
    folder icons/.style={folder, this folder size=#1, folder tree indent=3*#1},
    folder icons/.default={12pt},
}

\theorembodyfont{\upshape}
\theoremheaderfont{\scshape}
\theorempostheader{:}
\theoremsep{\newline}

\title[Results and Insights from Diagnostic Questions: The NeurIPS 2020 Education Challenge]{Results and Insights from Diagnostic Questions:\titlebreak The NeurIPS 2020 Education Challenge}

  \author{\Name{Zichao Wang}\footnote{\label{x}Equal Contribution.}\footnote{\label{rice}Rice University} \Email{jzwang@rice.edu}\\
  \Name{Angus Lamb}\footnoteref{x}\footnote{\label{msr}Microsoft Research Cambridge} \Email{t-anlam@microsoft.com}\\
  \Name{Evgeny Saveliev}\footnote{\label{uc}University of Cambridge} \Email{es583@cam.ac.uk}\\
  \Name{Pashmina Cameron}\footnoteref{msr} \Email{Pashmina.Cameron@microsoft.com}\\
  \Name{Yordan Zaykov}\footnoteref{msr} \Email{yordanz@microsoft.com}\\
  \Name{Jos\'{e} Miguel Hern\'{a}ndez-Lobato}\footnoteref{uc} \Email{jmh233@cam.ac.uk}\\
  \Name{Richard E. Turner}\footnoteref{uc} \Email{ret26@cam.ac.uk}\\
  \Name{Richard G. Baraniuk}\footnoteref{rice} \Email{richb@rice.edu}\\
  \Name{Craig Barton}\footnote{\label{eedi}Eedi} \Email{craig.barton@eedi.co.uk}\\
  \Name{Simon Peyton Jones}\footnoteref{msr} \Email{simonpj@microsoft.com}\\
  \Name{Simon Woodhead}\footnoteref{eedi} \Email{simon.woodhead@eedi.co.uk}\\
  \Name{Cheng Zhang}\footnoteref{msr} \Email{Cheng.Zhang@microsoft.com}\\
 }

\begin{document}

\maketitle

\editor{Editor's name}
\begin{abstract}
This competition concerns educational {\it diagnostic questions}, which are pedagogically effective, multiple-choice questions (MCQs) whose distractors embody misconceptions.  
With a large and ever-increasing number of such questions, it becomes overwhelming for teachers to know which questions are the best ones to use for their students.
We thus seek to answer the following question: how can we use data on hundreds of millions of answers to MCQs to drive automatic personalized learning in large-scale learning scenarios where manual personalization is infeasible?  
Success in using MCQ data at scale helps build more intelligent, personalized learning platforms that ultimately improve the quality of education en masse.
To this end, we introduce a new, large-scale, real-world dataset and formulate 4 data mining tasks on MCQs that mimic real learning scenarios and target various aspects of the above question in a competition setting at NeurIPS 2020.
We report on our NeurIPS competition in which nearly 400 teams submitted approximately 4000 submissions, with encouragingly diverse and effective approaches to each of our tasks.

\end{abstract}

\subsection*{Keywords}
Personalized education, Diagnostic questions, Question analytics, Unsupervised learning, Matrix completion, Missing value prediction, Active learning


\input{intro}

\input{data}

\input{tasks}

\input{lessons}

\bibliography{ref.bib}
\end{document}

%% file: intro.tex
\section{Introduction}
Recently years have seen an increasing proliferation of large-scale, online learning platforms that provide expertly produced materials and instructions at a low cost. These platforms are revolutionizing current education practices by lowering access to professional learning resources bringing high-quality learning experiences to the mass. One of the core enabling technologies of these platforms is personalized learning algorithms that automatically tailor instructions and pedagogical activities to each student considering their backgrounds, interests, and learning objectives.
However, \emph{personalization} remains a central challenge for these learning platforms and is still an active research area. This is because every student is unique and different, which requires individualized learning pathways that best suit each student.
While teachers, paying individual attention to each student, naturally adapt their pedagogy to the needs of that student, algorithms are less adaptable compared to expert teachers.

An over-arching question, then, is how to personalize an online learning platform, so that it adapts to the needs of a particular student.  That question is too big and vague, so in this paper, we focus on a small sub-problem: personalizing the choice of \emph{multiple-choice diagnostic questions}.
We chose this focus carefully.  First, there is ample research that shows that well-crafted multiple-choice questions are educationally effective~\citep{wylie2006dqs}. Second, multiple-choice questions make it easy to gather copious data in a very well-structured form: students, questions, and the answers students gave to some of those questions.
 Third, we had an active partnership with a live, deployed platform (Eedi) that had hundreds of thousands of users, so we could gather
 a \emph{lot} of data to feed today's data-hungry machine-learning algorithms.
 
 In this paper we describe a NeurIPS competition in which we supplied a large dataset of answers to such multiple-choice diagnostic questions and invited participants to engage in several tasks, all of which were directed towards the ultimate goal of identifying which questions would be most suitable for a particular student, at a particular point in their learning journey.  More specifically, our contributions are as follows:

\begin{itemize}
    \item We organized a competition at NeurIPS 2020 on {\it Multiple-Choice diagnostic Questions}, which we identify as a fertile application domain for machine learning in education.  Not only are these diagnostic questions educationally sound, but their format is also extremely well-suited to a range of machine learning methods.
    \item We introduce a massive dataset of answers to diagnostic questions, now publicly available (Section~\ref{sec:data}).\footnote{\label{ours-data}\url{https://eedi.com/projects/neurips-education-challenge}} This is by far one of the biggest datasets in the educational domain; see Table~\ref{dataset-compare} for a comparison. Our dataset also has the potential to help advance educational and machine learning research beyond the scope of our competition (Section~\ref{sec:impact}). 
    \item We introduce 4 different competition tasks that aim to address challenges relating to diagnostic questions. We include 2 common EDM tasks that our dataset enables, aiming at accurately predicting students' answers to questions (Sections~\ref{sec:task1} and~\ref{sec:task2}). More importantly, we also introduce a new task on automatic question quality assessment (Section~\ref{sec:task3}) and a task on personalized question selection (Section~\ref{sec:task4}). Our task design covers broader perspectives than previous competitions, e.g.,~\citep{Algebra,bridge}, which mostly focused on the first 2 tasks in our competition.
    \item We report on the results of our NeurIPS competition in which nearly 400 teams entered with almost 4000 submissions in total. We summarise the key insights from the leading solutions to each of the four tasks (Section~\ref{sec:results}), with discussions on the potential impacts of our competition, our dataset, and the submitted solutions on the future role of machine learning in education.
\end{itemize}

%% file: data.tex
\begin{table}[]
\centering
\caption{A Comparison of our new dataset to several existing ones. 
}
\vspace{-4pt}
\label{dataset-compare}
\scalebox{0.9}{
\begin{tabular}{@{}lrrr@{}}
\toprule
\textbf{Dataset} & \textbf{\#Students} & \textbf{\#Questions} & \textbf{\#Answer Records} \\ \midrule
Algebra2005\tablefootnote{\label{note1}\url{http://pslcdatashop.web.cmu.edu/KDDCup/downloads.jsp}}      & 569                 & 173,133              & 607,000                   \\
Bridge2006\footnoteref{note1}       & 1,135               & 129,263              & 1,817,427                 \\
ASSISTments2009\tablefootnote{\label{note2}\url{https://sites.google.com/site/assistmentsdata/home}}  & 4,151               & 16,891               & 325,637                   \\
Statistics2011\tablefootnote{\url{https://pslcdatashop.web.cmu.edu/DatasetInfo?datasetId=507}}   & 333                 & -                    & 189,297                   \\
ASSISTments2012\footnoteref{note2}  & 24,750              & 52,976               & 2,692,889                 \\
ASSISTments2015\footnoteref{note2}  & 19,840              & -                    & 683,801                   \\
ASSISTments2017\tablefootnote{\url{https://sites.google.com/view/assistmentsdatamining}}  & 1,709               & 3,162                & 942,816                   \\
NAEP2019\tablefootnote{\url{https://sites.google.com/view/dataminingcompetition2019/dataset}}         & 1,232               & 21                   & 438,291                   \\
\textbf{Ours}    & \textbf{118,971}    & \textbf{27,613}      & \textbf{15,867,850}       \\ \bottomrule
\end{tabular}}
\vspace{-5pt}
\end{table}

\section{Dataset}
\label{sec:data}

\begin{figure}[t]
    \centering
    \includegraphics[width=0.5\linewidth]{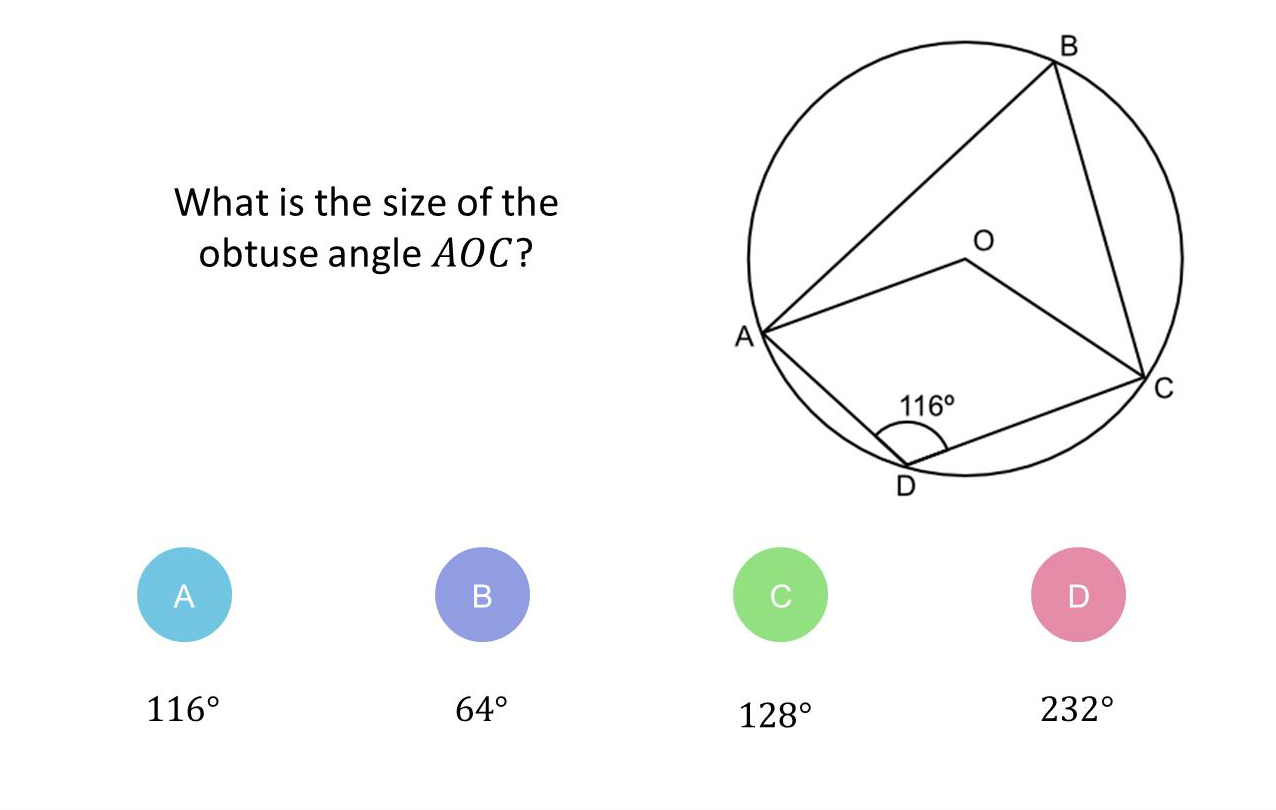}
    \vspace{-10pt}
    \caption{An example question in our dataset.}
    \label{fig:example_question}
    \vspace{-5pt}
\end{figure}

We curate a new, large-scale, real-world dataset from Eedi, an online education platform currently used in tens of thousands of schools, detailing student responses to multiple-choice diagnostic questions collected between September 2018 to May 2020. This platform offers crowd-sourced diagnostic questions to students from primary to high school (roughly between 7 and 18 years old). Each diagnostic question is a multiple-choice question with 4 possible answer choices, exactly one of which is correct. Currently, the platform mainly focuses on mathematics questions. Figure~\ref{fig:example_question} shows an example question from the platform. 

The competition is split into 4 tasks: tasks 1 and 2 share a dataset, as do tasks 3 and 4. These datasets are largely identical in format but use disjoint sets of questions. All QuestionIds, UserIds, and AnswerIds have been anonymized and have no discernable relation to those found in the product. Note that all such IDs for tasks 1 and 2 are anonymized separately from those for tasks 3 and 4, and the questions used in these two datasets are disjoint. This is by design, to ensure that the two datasets are both self-contained.

\subsection{Primary Data}

The primary training data for the tasks consists of records of answers that students responded to multiple-choice diagnostic questions. 
Table \ref{tab:exampledata} illustrates the format of some data points in our dataset. Because each student has typically answered only a small fraction of all possible questions, some students and questions are associated with too few answer records. Therefore, for tasks 1 and 2, we remove questions that have received fewer than 50 answers and students who have answered fewer than 50 questions. Similarly, for tasks 3 and 4, where we are interested in a fixed set of questions, we remove all students who had answered fewer than 50 of these questions. When a student has submitted multiple answers to the same question, we use the latest answer record. The data can be transformed into matrix form, where each row represents a student and each column represents a question.  Figure~\ref{fig:task1data} illustrates such a representation of our dataset.

For tasks 1 and 2, we randomly split the answer records into 80\%/10\%/10\% training/public test/private test sets. Similarly, for tasks 3 and 4, we randomly split the UserIds are into 80\%/10\%/10\% training/public test/private test sets. These pre-processing steps lead to training datasets of the following sizes:

\begin{itemize}
    \item Tasks 1 and 2: 27,613 questions, 11,8971 students, 15,867,850 answers
    \vspace{-5pt}
    \item Tasks 3 and 4: 948 questions, 4,918 students, 1,382,727 answers
\end{itemize}

The total number of answer records in these training sets exceeds 17 million, rendering manual analysis impractical and necessitating an automated, data-driven approach.

\subsection{Question Metadata}

We provide the following metadata for each question in the datasets.

\paragraph{SubjectId}  For each question, we provide a list of subjects associated with the question. Each subject covers an area of mathematics, at varying degrees of granularity. Example subjects include ``Algebra'', ``Data and Statistics'', and ``Geometry and Measure''. These subjects are arranged in a tree structure, i.e., the subject ``Factorising" is the parent subject of ``Factorising into a Single Bracket". 

 \paragraph{Question content.}
    In Tasks 3 and 4, in addition to the subjects, we also provide the image associated with each question, as shown in Figure \ref{fig:example_question}. Each image contains the question details including wording, figures, and tables.

\begin{table}[t]
\vspace{15.0pt}
\centering
\captionsetup{width=0.9\linewidth}
\vspace{-25pt}
\caption{Example answer records in our dataset. Each row represents one record.}
\vspace{-4pt}
\scalebox{0.9}{
\begin{tabular}{@{}lllccc@{}}
\toprule
{\bf QuestionId} & {\bf UserId} & {\bf AnswerId} & {\bf AnswerValue} & {\bf CorrectAnswer} & {\bf IsCorrect} \\ \midrule
10322 & 452 & 8466 & 4 & 4 & 1 \\
2955 & 11235 & 1592 & 3 & 2 & 0 \\
3287 & 18545 & 1411 & 1 & 0 & 0 \\
10322 & 13898 & 6950 & 2 & 1 & 0 \\
\bottomrule
\end{tabular}}
\label{tab:exampledata}
\vspace{-5pt}
\end{table}

\subsection{Student Metadata}

We provide the following metadata for each student in the datasets. 
\paragraph{UserId.} An ID that uniquely identifies a student. 
\paragraph{Gender.} A student's gender. 0 is unspecified, 1 is female, 2 is male and 3 is other.
\paragraph{DateOfBirth.} A student's date of birth, rounded to the first of the month. \paragraph{PremiumPupil.} A student's financial need status, i.e., a value of 1 indicates that a student is eligible for free school meals or pupil premium.

\subsection{Answer Metadata}

We provide the following metadata for each answer record in the dataset. \paragraph{UserId.} An ID that uniquely identifies an answer record.
\paragraph{DateAnswered.} The time and date that a student answered a question.
\paragraph{Confidence.} The percentage confidence score is given by a student when answering a question. 0 means a random guess, 100 means complete confidence.
\paragraph{GroupId.} The class (a group of students) in which a student was assigned a question.
\paragraph{QuizId.} The assigned quiz which contains the question the student answered. \paragraph{SchemeOfWorkId.} The scheme of work in which the student was assigned the question. A scheme of work is a sequence of topics which contain quizzes. A scheme of work typically lasts for one academic year.

%% file: tasks.tex
\section{Competition Tasks} \label{sec:tasks}

\noindent
In this section, we describe the 4 competition tasks and the evaluation metrics.
The first two tasks aim to predict the student's responses to every question in the dataset. These two tasks can be formulated in several ways, for instance as a recommender system challenge \citep{harper2015movielens,bennett2007netflix,dror2012yahoo} or as a missing value imputation challenge \citep{little2019statistical, yoon2018gain, stekhoven2012missforest, gong2019icebreaker}. 
The third task focuses on evaluating question quality which is essential and remains an open question in the education domain \citep{wang2020large}. 
The fourth task directly addresses the challenge of personalized education where personalized dynamic decision making \citep{agarwal2016multiworld, li2010contextual, eddi} is needed.

\begin{figure}[t]
    \centering

    \includegraphics[width=0.47\linewidth]{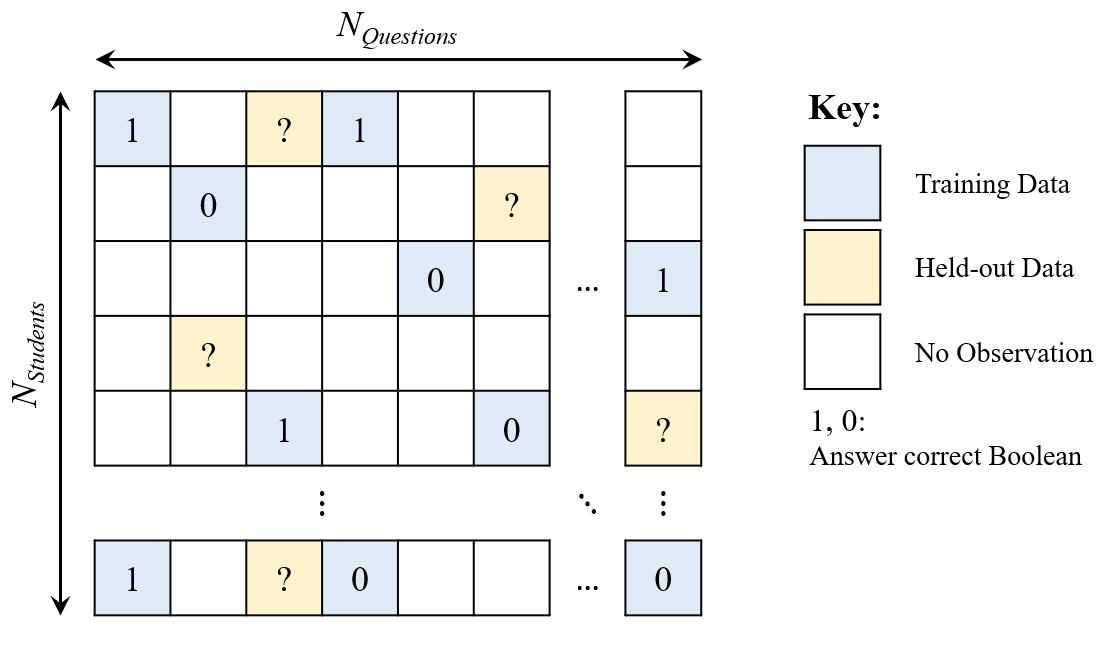}
    \hspace{10pt}
    \includegraphics[width=0.47\linewidth]{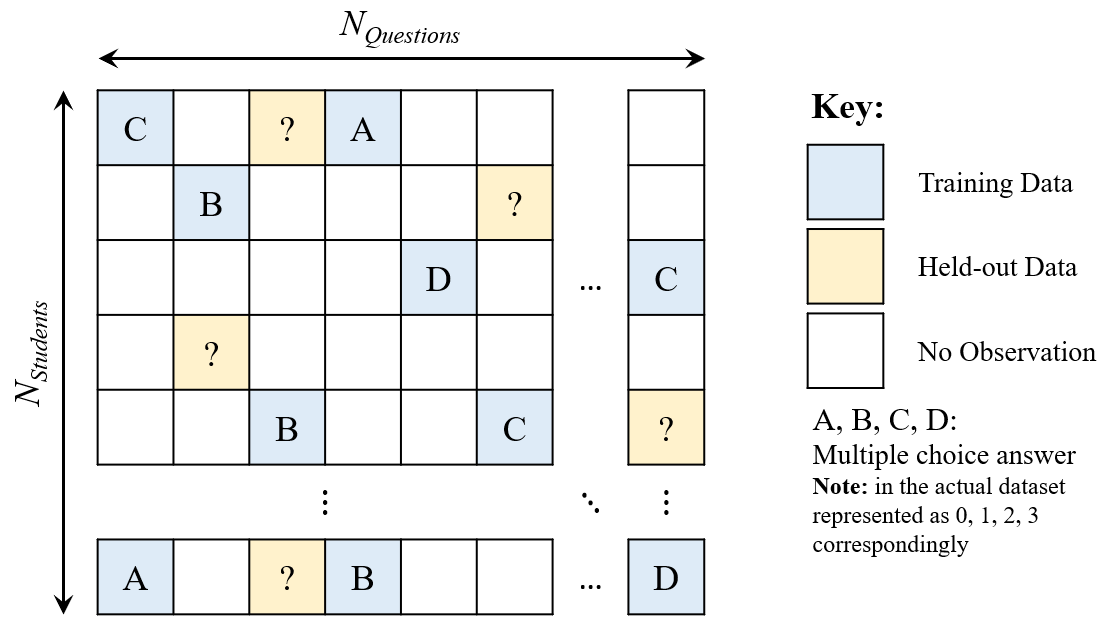}
    \vspace{-5pt}
    \caption{Illustrations of the sparse matrix representation of the data for Task 1 ({\bf left}) and Task 2 ({\bf right}). 
    }
    \label{fig:task1data}
    \vspace{-5pt}
\end{figure}

\subsection{Task 1: Predict Student Responses -- Right or Wrong} 
\label{sec:task1}
The first task is to predict whether a student answers a question correctly. The primary data used for this task is a table of records (\texttt{StudentId}, \texttt{QuestionId}, \texttt{IsCorrect}) where the last column is a binary indicator for whether the student answered the question correctly. A sparse matrix representation of this data is illustrated in Figure \ref{fig:task1data}. Specifically, for each student, a portion of the available records was held out as the hidden test set on which evaluation was performed.

\paragraph{Evaluation Metric.}
We use {\bf prediction accuracy} as the metric, i.e., the number of predictions that match the true correctness indicator, divided by the total number of predictions (in the held-out test set): 
\begin{align*}
    {\rm Accuracy} = \frac{\#{\rm correct\,predictions}}{\#{\rm total\,predictions}}
\end{align*}

\paragraph{Significance.} Predicting the correctness of a student's answers to unanswered (or newly introduced) questions is crucial for estimating the student's skill levels in a real-world personalized education platform and forms the basis for more advanced tasks. This task falls under the class of matrix completion and is reminiscent of challenges often seen in the recommender systems domain in the case of binary data. 

\subsection{Task 2: Predict Student Responses --  Answer Prediction:} 
\label{sec:task2}
The second task is to predict which answer a student responds to a particular question. The primary data used for this task is a table of records (\texttt{StudentId}, \texttt{QuestionId}, \texttt{AnswerValue}, \texttt{CorrectAnswer}) where the last 2 columns are categorical taking values in [1, 2, 3, 4] (corresponding respectively to multiple-choice answer options A, B, C and D). The sparse matrix representation is illustrated in Figure \ref{fig:task1data} (right). Because the questions in our dataset are all multiple-choice, each with 4 potential choices and 1 correct choice, we treat this task as a multi-class prediction problem in a matrix completion formulation. This problem formulation is similar to that in Task 1 but with unordered categorical data, i.e., students' actual choices, instead of binary data, i.e., students' correct/incorrect answer indicators. We note that such unordered, categorical data is rare in the recommender systems domain, where responses will typically be binary or ordinal (e.g. 1-5 stars). 

\paragraph{Evaluation Metric.}
We use the same metric {\bf prediction accuracy} as in Task 1, except that the true answers are now categorical instead of binary.

\paragraph{Significance.} Predicting the actual multiple-choice option for a student's answer allows analysis of likely common misconceptions that a student may hold on a topic. For example, clusters of question-answer pairs that are highly correlated may indicate that they correspond to the same, or related misconceptions. Understanding the relationships among misconceptions is a crucial problem to solve for curriculum development, which could inform the way a topic is taught and the sequencing of topics.

\noindent

\subsection{Task 3: Global Question Quality Assessment} 
\label{sec:task3}
The third task is to predict the ``quality" of a question, as defined by a panel of domain experts (experienced teachers), based on the information learned from the students' answer records. Because how expert teachers judge question quality is unknown and difficult to quantify, this task requires defining and developing a metric for evaluating the question quality that mimics the experts' judgment of question quality. 
This task can be viewed as an unsupervised learning task because there is no explicit supervision label available for question quality.

\begin{figure}[t]
    \centering
    \includegraphics[width=0.65\linewidth]{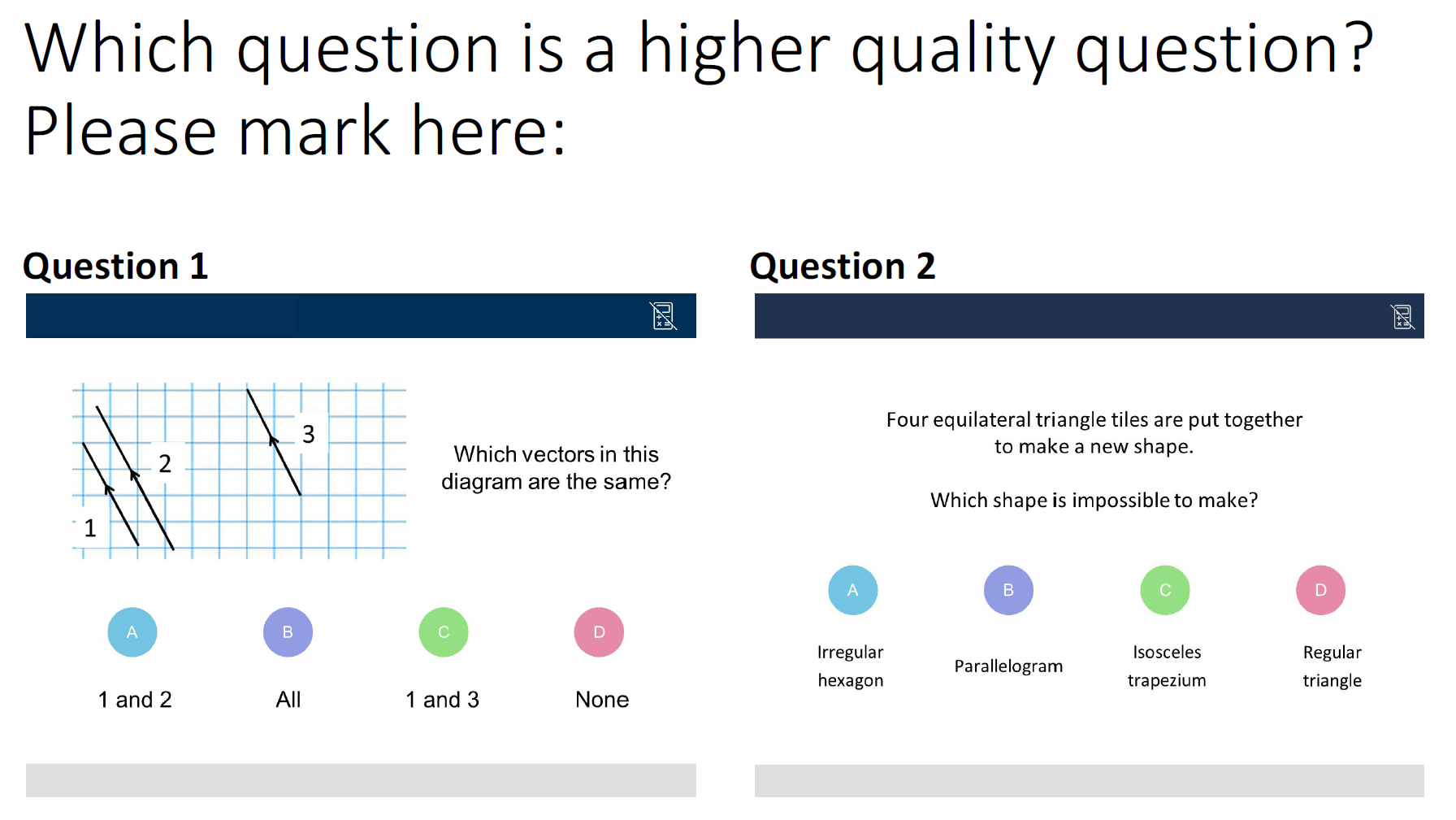}
    \vspace{-5pt}
    \caption{Example of a prompt used in collecting experts' judgement of pairwise relative question quality. In addition to this, the experts receive the following instructions: \textit{On each of the following slides, you will see 2 questions, one on the left and one on the right. Please decide which question is of higher quality; ties are not allowed.}}
    \label{fig:task3expert}
    \vspace{-5pt}
\end{figure}
\begin{figure}
    \centering
    \includegraphics[width=0.9\linewidth]{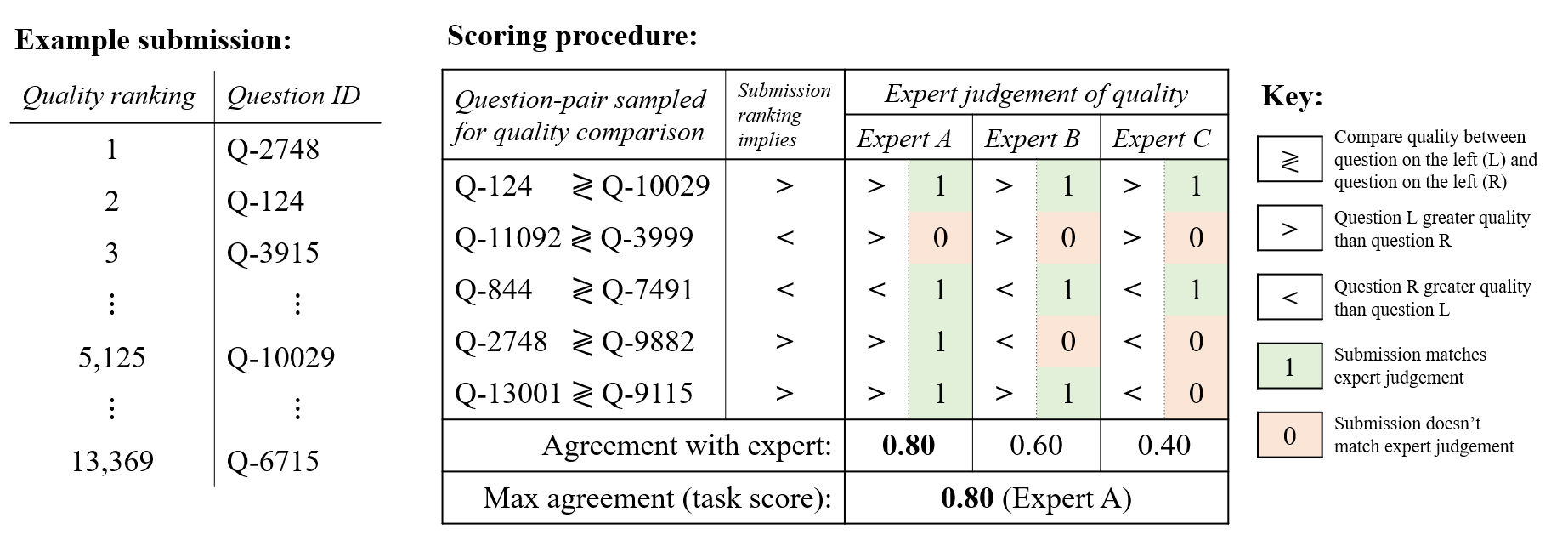}
    \vspace{-5pt}
    \caption{An illustration of the scoring process in Task 3. \textit{Left}: The expected format of the submissions for Task 3 - a ranking of question quality over the \texttt{Question ID}s, in the decreasing order of quality. \textit{Right}: An illustration of the performance metric calculation; see Section \ref{task3eval} for more details. This example uses 5 question-pairs and 3 experts.}
    \label{fig:task3metric}
    \vspace{-5pt}
\end{figure}

\paragraph{Evaluation Metric.} \label{task3eval}
We evaluate the agreement between the automatically computed question quality ranking using the proposed quality metric with the experts' rankings. 
To gather data for evaluation, we manually collect pairwise question quality rankings on a subset of questions from 5 different expert evaluators.
An example of a prompt used in the data collection process is shown in Figure \ref{fig:task3expert}. In addition, a few ``Golden rules" of quality question design have been identified by one of the domain experts, Craig Barton,\footnote{\url{https://medium.com/eedi/what-makes-a-good-diagnostic-question-b760a65e0320}} which guides the quality ranking. Specifically, high-quality questions should 
\begin{itemize}
\item be clear and unambiguous;
\vspace{-5pt}
\item test a single skill/concept;
\vspace{-5pt}
\item allow students to be able to answer them in less than 10 seconds;
\vspace{-5pt}
\item have carefully designed answer choices such that one can easily identify the misconceptions from each incorrect answer choices;
\vspace{-5pt}
\item be challenging to answer for students with key misconceptions that the questions intend to identify.
\end{itemize}
In total, we randomly select 40 pairs of questions and ask each of our evaluators to separately provide a binary ranking to each pair, i.e., which question in the pair is of higher quality.
The evaluation steps are then as follows (see also Figure \ref{fig:task3metric} for an illustration). First, a question quality metric compute a full ranking for all questions in descending quality. Second, we extract the pairwise ranking of the 40 pairs of questions from this full ranking. Third, for each expert $i$, we determine the agreement fraction: $A_i = \frac{N_{matching\_pairs}}{N_{total\_pairs}}$. Lastly, we find the \textit{maximum} of these agreement fractions $A_{max} = \max_{i} A_i$. This $A_{max}$ score serves as the final evaluation metric for this task.
    
We were looking for metrics that can approximate any expert's judgment really well, hence we used the maximum of the agreement fractions, rather than a mean of the agreement fractions over all experts. The reasoning for this approach is that the quality metrics of the experts are in themselves subjective, and it is interesting to find whether a particular expert's approach can be approximated especially well by the use of machine learning.

\paragraph{Significance of Task 3.} A scalable, evidence-based mechanism that produced reliable measures of question quality remains an open challenge because question quality is often regarded as a subjective measure, i.e., different people might have varying definitions of question quality. Thus, having one unified, an objective metric is challenging. 
Furthermore, even experts sometimes write questions that seem good at first glance but turn out to be of poor quality. In situations where questions are crowd-sourced, the question quantity is large but the quality is likely to vary significantly, rendering manual inspection challenging and desiring automated techniques to identify high-quality questions. Automatic question quality judgment can also be a valuable guide for the teachers, i.e., to help them author higher quality questions.

\subsection{Task 4: Personalized Questions} 
\label{sec:task4}
\begin{figure}[t]
    \centering
    \includegraphics[width=0.65\linewidth]{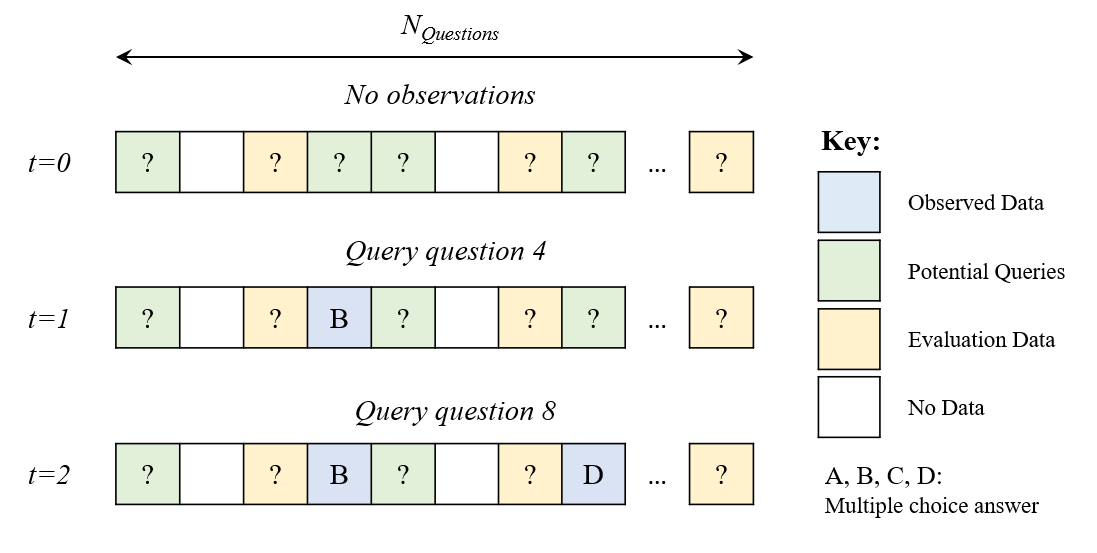}
    \vspace{-5pt}
    \caption{An illustration of the procedure for Task 4. On each time step, the model can train on the data in blue, and its predictive performance is assessed on the held-out data in yellow. The algorithm must then choose the next question to query from the set of green questions using this new model.
    }
    \label{fig:task4data}
    \vspace{-5pt}
\end{figure}

The fourth task is to interactively generate a \emph{sequence} of questions to ask a student to maximize the predictive accuracy of a model on their remaining answers. Specifically, a participant's model was provided with a set of previously-unseen students, whose answers to questions were completely hidden, and a set of potential questions to query for each student. 
This task can be viewed through the lens of several related fields, including active learning, reinforcement learning, bandit algorithms, Bayesian experimental design, and Bayesian optimization, and insights drawn from any of these fields will likely prove useful. Figure~\ref{fig:task4data} illustrates the setup of this task.

\paragraph{Evaluation Metric.}
Submitted models were asked to sequentially choose 10 query questions for every student in a held-out set of students. After each selection step,  both the categorical answer and binary correctness indicator for these student-question pairs were revealed to the model in private. The model was then allowed to incorporate this new data or retrain after each question. After receiving 10 answers for each student, the model was assessed on its prediction accuracy for predicting the binary correctness for a held-out test set of answers for each of these students whose answers cannot be queried.

\paragraph{Significance of Task 4.} the task aims to maximize the predictive accuracy of a participant's model on a held-out set of questions for each student after the model has been exposed to the 10 answers from each student. This task is of fundamental importance to personalized education, where we wish to accurately diagnose a student's level of understanding of various concepts while asking the minimum number of questions possible, to make the most efficient use of both student and teacher time. The task is also a crucial machine learning challenge, requiring effectively reasoning about a machine learning model's uncertainty and efficiently using data.

%% file: lessons.tex
\section{Competition Results} \label{sec:results}

Over the course of 2 months from July 2020 to October 2020, Our competition attracted in total 3,696 submissions from 382 teams worldwide. Across all 4 tasks, we observe significant improvements over time. Table~\ref{tab:submission_stat} shows the participation and submission breakdown for each task. 
In the following sections, we briefly describe the top team's solution for each task. We then summarize our insights and findings from the competition.

\subsection{Winning solutions}

\paragraph{Top solution for Task 1.}

The winning solution~\citep{takehara2020practical} comes from Aidemy, Inc., which achieves 77.29\% prediction accuracy. Their solution takes advantage of useful feature engineering, feature selection, and ensemble methods. The primary technique that they use involves {\em target encoding}, which computes the target statistics (in this case, the students' binary correct/incorrect answer records) based on segments of features, i.e., split the data based on students' demographics and compute key statistics on each subset. The authors use this technique in many of their solutions and demonstrate its effectiveness in this task in particular. 

In addition, they construct several features from various {\em metadata}, three of which they find particularly helpful in improving the prediction performance. The first type of feature is {\em time-related} features, such as the timestamps of the students' answers and students' date of birth (age). The second type of feature is {\em user history} features which include students' experience to date, such as questions answered and the number of classes taken. The third type of feature is {\em subject} features including question subjects. Some additional useful features include vectors obtained by taking singular value decomposition (SVD) on the student-question answer matrix, which reveals important student-question interaction patterns. Besides using these features as input to the prediction model, they also used them together with target encoding, i.e., to group students into different age groups or group questions based on subjects.

The prediction pipeline goes through three steps. The first {\em feature selection} step uses an ensemble of models, each selecting the top 100 features. The second {\em meta feature construction} step uses another ensemble of models, is trained on the selected features, and produces meta-features. The final step uses the meta-features as input to train various regression models, again in an ensemble manner, to produce the final prediction. 
\begin{table}[]
\centering
\caption{Number of active teams and total submissions for each task. 
}
\label{tab:submission_stat}
\vspace{-4pt}
\scalebox{0.9}{
\begin{tabular}{@{}lcccc@{}}
\toprule
Tasks          & Task 1 & Task 2 & Task 3 & Task 4 \\ \midrule
\#Teams & 80 & 33 & 16 & 17 \\
\#Submissions & 1401   & 448    & 1061   & 786    \\ \bottomrule
\end{tabular}}
\vspace{-5pt}
\end{table}
\paragraph{Top solution for Task 2.}
The winning solution~\citep{shen2020which} comes from a team at the University of Science and Technology of China, which achieves 68.03\% prediction accuracy. In their solution, they proposed a novel {\em order-aware cognitive diagnosis model} (OCD), which is inspired by the idea to capture students' {\em attention span} when answering questions. They implement a convolutional neural network and use the sliding convolutional windows to capture the students' attention span. Specifically, they first organize students' answer records as a list of tuples $\{(q_i,c_i)\}_i$ where $q_i$ and $c_i$ are the question and student embedding, respectively. Second, they concatenate all records of one student into a single vector. Lastly, they apply a stacked 1-dimensional sliding convolutional neural network (CNN) to convert each student record into a feature. This implementation is ``order-aware'' because the students' answer records are arranged in order of their answer timestamps. In addition, they use an ensemble of CNNs with different sliding window sizes to model multi-scale attention span. In their experiments, they empirically demonstrate the advantage of modeling multi-scale attention spans versus modeling only a single attention span.

\paragraph{Top solutions for Task 3.}
The winning solutions come from three different teams~\citep{shinahara2020quality,tal2020solution,mcbroom2020assessing}, all of which achieve a maximum of 80\% agreement with human evaluators' judgments. This is because the winning criterion takes the maximum agreement with any of the evaluators. In table~\ref{tab:task3-detail} we show the full results, i.e., the agreement between each of the evaluators and each winning team's submission. Below, we briefly describe each of the three winning solutions. 

The team from Aidemy, Inc. \citep{shinahara2020quality} presents a solution based on the hypothesis that a high-quality question is appropriately difficult, is readable, and strikes a balance among answer choices. They compute a question feature for each of the hypothesized properties. Specifically, they use entropy to measure question difficulty and the balance among answer choices. They also extract readability features by performing optic character recognition (OCR) on the question images to extract question textual information. Their empirical results show that combining all the above features results in the best performance on both public and private data compared to using only a subset of the features. To compute the question quality rankings, they first rank the questions using each feature and then rerank them by the average rank. 

The team from TAL Education Group \citep{tal2020solution} presents a solution that shares some similarity with the above solution. Specifically, they also compute entropy among answer choices and entropy between correct and incorrect student responses. Here, they additionally compute fine-grained correct-incorrect response entropy conditioned on question groups and quiz groups, which captures question difficulty at different levels of knowledge acquisition. They also include students' answer confidence scores into their metric. Different from the above solution, they form their metric by taking a weighted average of the features to compute the final ranking. 

The team from the University of Sydney \citep{mcbroom2020assessing} provided an even simpler solution, simply ranking the questions in order of the average confidence students reported in their answers to each question, where higher average confidence is interpreted as a higher question quality. The authors propose that high student confidence implies that the question is clear and unambiguous and that the Dunning-Kruger effect \citep{kruger1999unskilled} may result in students holding key misconceptions reporting high confidence in incorrect answers if the question clearly addresses this misconception.

Notably, all of these solutions are entirely deterministic and do not involve any machine learning component.

\begin{table}[]
\centering
\caption{Detailed results of winning solutions to Task 3, including the agreement with each of the evaluators.}
\label{tab:task3-detail}
\vspace{-4pt}
\scalebox{0.9}{
\begin{tabular}{@{}lccccc@{}}
\toprule
\textbf{Team} & \textbf{Eval \#1} & \textbf{Eval \#2} & \textbf{Eval \#3} & \textbf{Eval \#4} & \textbf{Eval \#5} \\ \midrule
Aidemy, Inc   & 0.64              & 0.6               & 0.6               & 0.6               & 0.8               \\
TAL           & 0.72              & 0.68              & 0.6               & 0.6               & 0.8               \\
U. Sydney     & 0.72              & 0.6               & 0.6               & 0.6               & 0.8               \\ \bottomrule
\end{tabular}}
\vspace{-5pt}
\end{table}
\paragraph{Top solution for Task 4.}
The winning solution comes from Aritra Ghosh at the University of Massachusetts, Amherst \citep{ghosh2020meta}, who proposes a novel {\em meta-learning} framework. The intuition is that a model needs to quickly adapt to each student, i.e., it selects a personalized sequence of questions for each student, after observing only a few answer records from each student. This few-shot learning setup during test time implies that it is best if the training procedure also follows this few-shot learning setup. Therefore, they formulate a meta-learning problem, where the model aims to first optimize different small subset of answers, partitioned from the training data, for each student and then optimize another subset of answers distinct from the training data. As part of their solution, they also propose a sampling-based method to select the next question conditioned on the already observed answers. They note that a more complicated, potentially more powerful question selection policy, such as the actor-critic network in reinforcement learning, can be incorporated into their framework. 

During training, they use different partitions of the training data to train the model. Several other training and data splitting techniques are involved, such as splitting all students into training and local test sets in addition to splitting answer records for each student. These techniques together ensure that the model does not overfit and is capable of generalizing to different students with very few initial answer records. 

\section{Observations and Insights}

\paragraph{The importance of educational domain knowledge.}
The successful solutions in all tasks demonstrate the importance of educational domain knowledge. For example, the best solution for task 1 leverages several related metadata about students and questions and combines them with feature selection and ensemble methods. This observation suggests that to build successful predictive and analytical models, it is important to creatively and cleverly incorporate educational domain knowledge into potentially black-box machine learning methods, in addition to simply building more powerful models. Empirically, we observe that methods that purely rely on state-of-the-art models such as deep learning perform sub-par with those that take advantage of educational domain knowledge and intuitions. Models with domain knowledge built-in also have the potential to provide us valuable insights into which features are useful for the predictive task, which can then help guide and optimize the data collection process.  

\paragraph{The potential utility of entropy-based question quality metrics.}
Many of the top-performing solutions leverage entropy as a way to measure question quality. 
Entropy is an appropriate measure of balance among a question's answer choices and between the correct and incorrect students' responses. 
As some submitted solutions' intuition indicates, questions that strike such balance tend to be moderately difficult (not too easy or too difficult) and can distinguish students that have or do not have a mastery of knowledge. 
Those questions thus should be of high quality.

The above observations yield two interesting insights. First, the aforementioned intuitions closely align with the expert human evaluators' judgment criteria, because question quality ranked by metrics based on those intuitions achieve substantial agreements with that ranked by expert human evaluators. Second, entropy is a suitable way to quantify the above intuition. Our insights suggest that entropy-based metrics are a promising direction to explore for objectively quantifying question quality, a property that we traditionally consider as overly subjective and unlikely to be computed. 

\paragraph{The promise of emerging ML techniques for education.}
Across all submissions, we have seen many creative applications of the latest machine learning methods.
For example, the winning solution in task 4 develops a meta-learning framework that also leverages techniques such as reinforcement learning, few-shot learning, and bi-level optimization. 
Although these novel methods sometimes fall behind traditionally successful methods in data science competitions such as boosting methods and ensembles, they bring in new ideas to modeling educational data and have the potential to contribute to other practical data science problems beyond education.

\section{Competition Impacts}
\label{sec:impact}
In this section, we describe the potential impacts that our competition has on AI for education. We also discuss other educational and broader impacts.

\paragraph{Impact on AI for Education.}
As described in Section~\ref{sec:tasks}, each of the competition tasks is rooted in a genuine, real-world educational problem. Successful solutions to tasks 1 and 2 will lead to more accurate student analytics and misconception identification, respectively. These improvements could better assist teachers and personalized learning algorithms in knowing how the students are learning, leading to more effective personalized learning. Successful solutions to task 3 provide novel ideas to quantify question quality. These ideas will spark new research in question quality quantification and provide preliminary ways to do so.
Finally, successful solutions to task 4 will lead to more effective question sequence selection. This will potentially improve performance for adaptive testing algorithms, improve efficiency for online assessment, and save teachers' time in manually selecting questions.

Our competition may potentially benefit AI for education beyond the tasks that we introduced. Our dataset, which is large in size and rich in metadata, will contribute to many other research problems in AI for education. For example, our dataset contains the timestamps of the answers, which fits perfectly in the setting of knowledge tracing~\citep{kt,pkt,dkt,akt}, one of the fundamental problems in educational data mining that tracks students' progress over time. Our dataset also contains the topics/skills that each question intends to test, which can be used for fine-grained misconception identification and analysis~\citep{mc1,mc2,mc3,mc4,mc5}. Finally, our dataset contains question images, which include the question text.  These texts and images could enable research on multi-modal data integration, i.e., images and natural language, to improve modeling performance on a variety of educational data mining tasks.

\paragraph{Broader Impacts.}

Our competition also involves multiple fundamental machine learning challenges that need to be addressed. Some challenges are common in recommender systems but appear in the context of educational data mining: how to deal with the sparsity of the data because each student answered only a small fraction of all questions? How to effectively use student and question metadata, such as student demographics, to improve the prediction? How to account for the fact that the students themselves are not static, but are learning through time? Other challenges are reminiscent of active learning: how to optimally select the sequence of questions to maximize prediction accuracy? Another challenge is how to effectively perform matrix completion for unordered, categorical data. Tackling these challenges in the unique educational context will be of significant technical interest to the NeurIPS community and the machine learning community as a whole. 

\section{Conclusions and Future Work}
We present the NeurIPS 2020 Education Challenge, a comprehensive challenge focusing specifically on the applications of AI in education. We introduce 4 different tasks, all of which represent practical and pressing problems that large-scale online educational platforms are facing today. We curate one of the largest educational datasets available to date with rich metadata on both students and questions. The competition has attracted wide attention from participants worldwide, whose solutions bring fresh ideas to large-scale educational data mining and suggest promising future research directions for each of the tasks. Finally, we note that our large, real-world dataset has much broader applications in both AI for education and machine learning than those tasks that we designate in our competition. 

Our competition deliberately pinpoints a focused target: student responses to multiple-choice diagnostic questions.
But education is extremely complicated, and there are many, many other important problems. For example, it is difficult to define and quantify students' learning outcomes. In our current setting, students' objective is to correctly answer all questions assigned to them. This may not be a poor approximation of the real learning outcome that teachers care about. Also, currently, we can only analyze students' performance {\it after} we collect their data. It is highly desirable, but challenging, to perform analysis, and even {\it intervention} during students' learning process, but this may yield ethical issues. Besides, many other types of learning activity data such as collaboration, emotional state, and confidence, which potentially have a high impact on learning, are difficult to collect. Finally, it is an ongoing research problem to determine how best to collect and save large educational data for AI applications while preserving students' and teachers' privacy. We are continuing to work on these challenging issues and we believe that our dataset and the insights drawn from the competition will have a positive long-term impact on educational practitioners and diverse research communities. 